# Science and Technology Progress at the Sydney University Stellar Interferometer

J. Gordon Robertson*[a], Michael J. Ireland[b,c], William J. Tango[a], Peter G. Tuthill[a], Benjamin A. Warrington[b], Yitping Kok[a] , Aaron C. Rizzuto[b], Anthony Cheetham[a], and Andrew P. Jacob[a]

[a]Sydney Institute for Astronomy, School of Physics, University of Sydney, NSW 2006, Australia;
[b] Department of Physics and Astronomy, Macquarie University, NSW 2109, Australia;
[c] Australian Astronomical Observatory, PO Box 296, Epping, NSW 2121, Australia

## ABSTRACT

This paper presents an overview of recent progress at the Sydney University Stellar Interferometer (SUSI). Development of the third-generation PAVO beam combiner has continued. The MUSCA beam combiner for high-precision differential astrometry using visible light phase referencing is under active development and will be the subject of a separate paper. Because SUSI was one of the pioneering interferometric instruments, some of its original systems are old and have become difficult to maintain. We are undertaking a campaign of modernization of systems: (1) an upgrade of the Optical Path Length Compensator IR laser metrology counter electronics from a custom system which uses an obsolete single-board computer to a modern one based on an FPGA interfaced to a Linux computer - in addition to improving maintainability, this upgrade should allow smoother motion and higher carriage speeds; (2) the replacement of the aged single-board computer local controllers for the siderostats and the longitudinal dispersion compensator has been completed; (3) the large beam reducing telescope has been replaced with a pair of smaller units with separate accessible foci. Examples of scientific results are also included.

## 1. INTRODUCTION

Stellar interferometry through the Earth's turbulent atmosphere is a challenging task, but provides direct measurements of stellar angular diameters and the orbits of close binary stars. After the early work of Michelson and Pease[1] using two apertures along a 20-foot beam atop the Mt Wilson 100 inch telescope, and the later development of intensity interferometry[2], one of the first 'modern Michelson' instruments was the Sydney University Stellar Interferometer (SUSI)[3,4].

SUSI is a 2-element interferometer with North-South baselines selectable from 5m to 160m. Table 1 gives the basic parameters of the instrument and Figure 1 shows the overall optical paths up to the beam combiners. As one of the pioneering instruments, SUSI has been through several generations of detector technology. Originally set up with the blue table beam combiner[4] using photomultiplier detectors, SUSI was later converted to use Avalanche Photo Diodes as fringe detectors and a CCD for the tip-tilt detector in the red table beam combiner[5]. More recently, advantage has been taken of the availability of fast-readout low-noise electron-multiplying CCDs, which enable a 2-dimensional readout. The PAVO beam combiner now uses such a detector to provide wavelength-dispersed fringe detection, as well as tip-tilt signals.

*Gordon.Robertson@sydney.edu.au; phone +61 2 9351 2825; fax +61 2 9351 7726

Table 1. Summary of SUSI specifications

| Location | |
|---|---|
| Latitude | 30°19' 20.186" S |
| Longitude | 149° 32' 53.610" E |
| Altitude | 211m |
| Orientation | North - South |
| Baselines | 5-160 m |
| **Operating Wavelengths** | |
| Fringe-tracking (PAVO) | 530 – 810 nm |
| Astrometry (MUSCA) | > 810 nm |
| Tip-tilt | < 530 nm |
| Metrology (OPLC) | 1.152591μm |
| Metrology (MUSCA) | 543.3 & 632.8 nm |

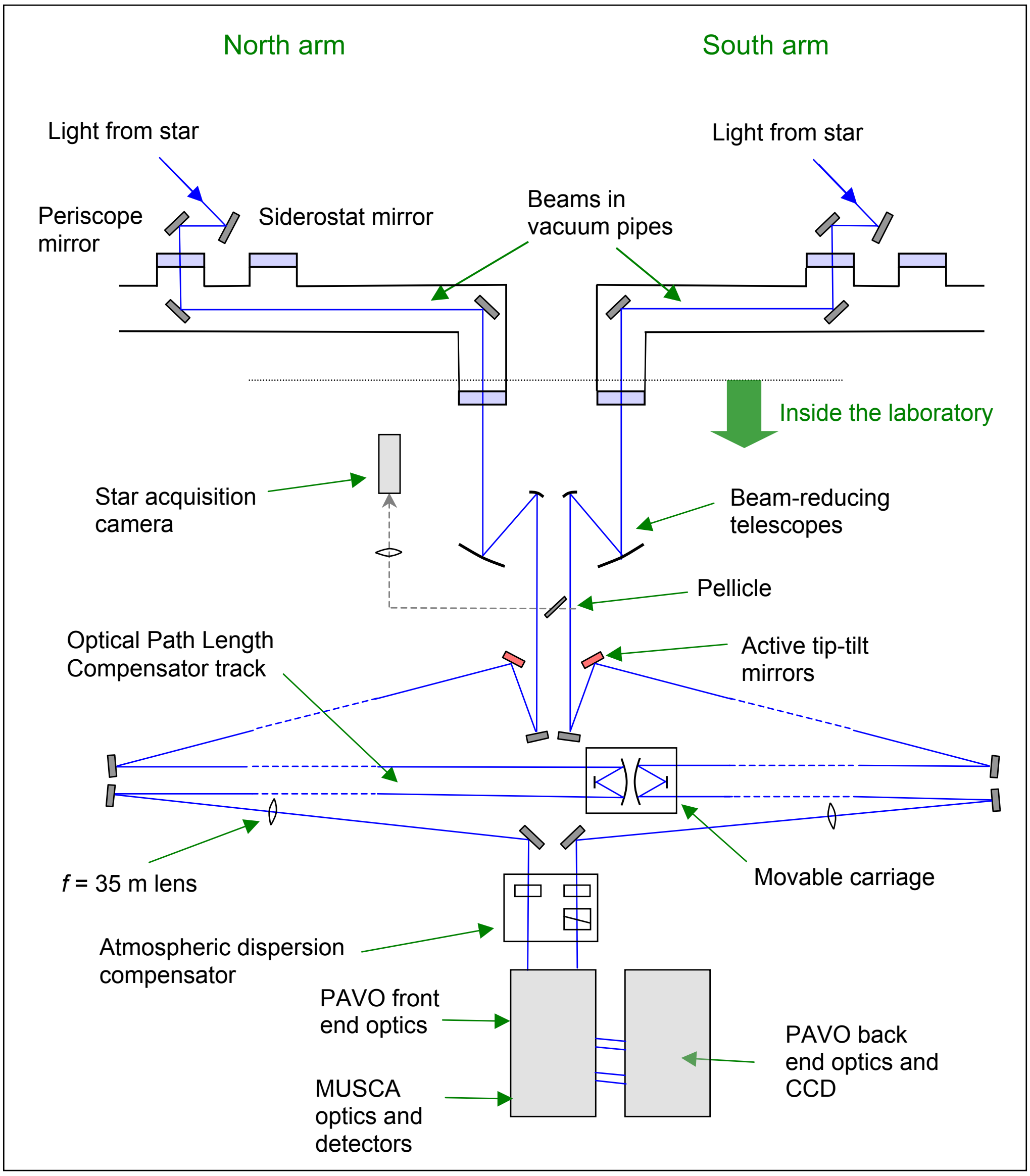

Figure 1. Schematic outline of the optical paths in SUSI.

# 2. THE PAVO BEAM COMBINER – CURRENT STATUS

The Precision Astronomical Visible Observations (PAVO) beam combiner concept[6] uses an electron-multiplying (Low Light Level) CCD, which is a 2-dimensional detector with sufficiently fast readout to handle the rapid sampling required to 'freeze' atmospheric turbulence effects. PAVO uses one dimension of the detector to sample the fringes and the other dimension on the detector allows wavelength dispersion of the fringes. This provides a number of independent wavelength channels which increases sensitivity and information content, but does not require the full detector width, so the instrument's pupil can be segmented and fringes registered separately from each segment. This ameliorates the reduction in fringe visibility due to seeing and any optical aberrations in the system. Further areas of the same detector are used to receive the tip-tilt sensing images.

The 20 cm siderostats shown in Figure 1 send 14 cm beams down vacuum pipes to the central laboratory, then through a 3:1 beam reduction telescope, via piezo-actuated tip-tilt mirrors to a delay line carriage incorporating piezo-actuated secondary mirrors[4].

One of the innovations in PAVO is the use in each beam of a weak positive lens with a focal length of ~35m situated on the OPLC plinth at 35 m from the PAVO central pier optics. This lens results in very small beam diameters at the central pier and PAVO backend optics, enabling the use of small diameter optics and a knife-edge mirror.

The beams pass through the longitudinal dispersion corrector (LDC[4]) and then enter the PAVO central pier optics as shown in Figure 2. At this stage the beam diameter is only a few mm.

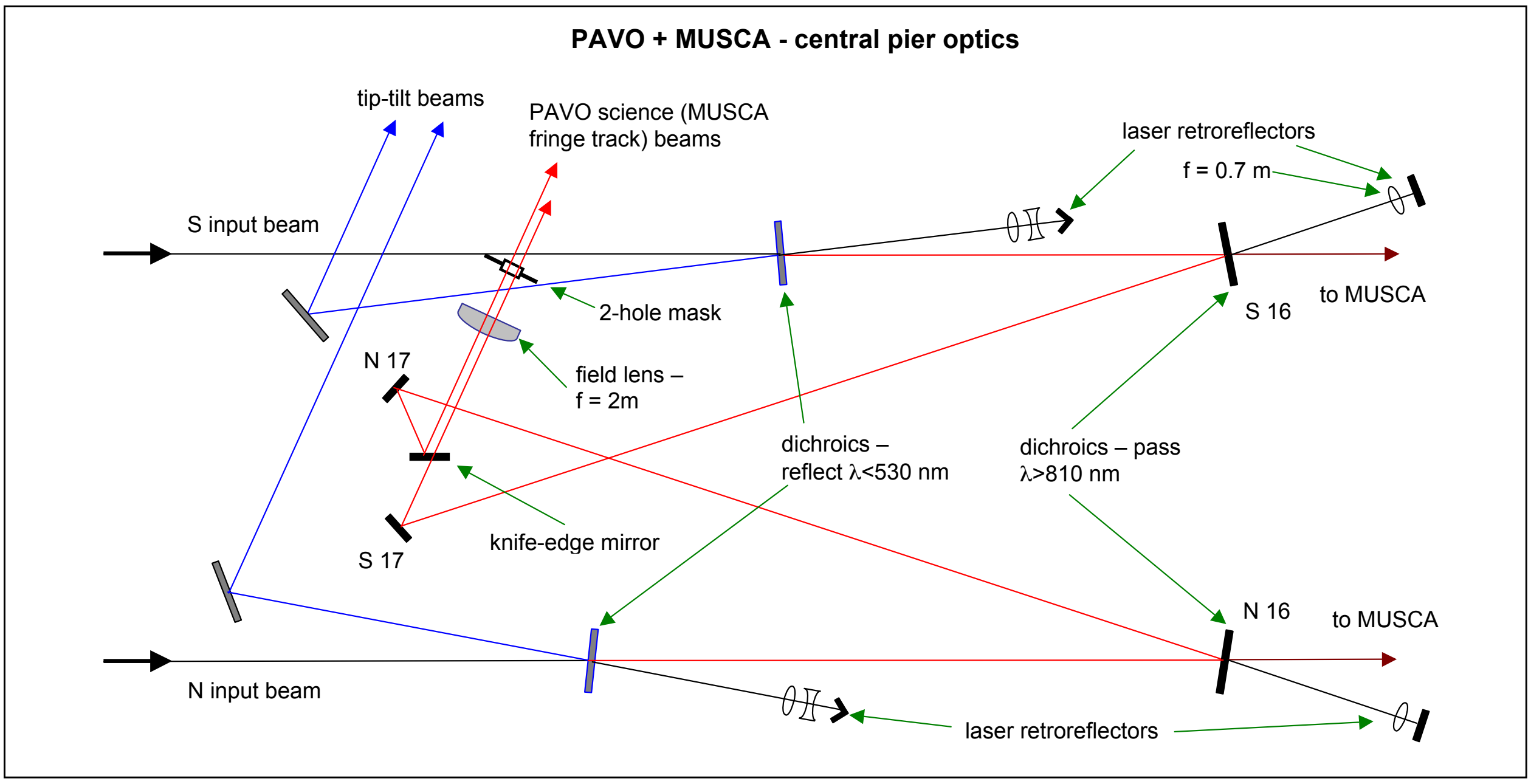

Figure 2. Schematic diagram of the light path and the components on the central pier. Laser retroreflectors are used in conjunction with the lasers shown in Figure 3, to achieve low-level laser injection into the MUSCA data frames. They rely on the fact that a small fraction of the laser beam is transmitted through a dichroic although that wavelength is nominally reflected.

Within the central pier optics, dichroics partition the beams into three parts (see Table 1): the shortest wavelengths go to the tip-tilt sensor windows of the CCD, the intermediate band goes to PAVO for fringe detection, and the longest wavelengths go to the MUSCA differential astrometry instrument. When operated in conjunction with MUSCA (see Section 3) the role of the PAVO 'science' beams will be to provide accurate fringe tracking.

Figure 3 shows the back-end optics. The two tip-tilt beams and the two science beams pass through a slit baffle and then on to an array of vertical-axis cylindrical lenslets with focal length 2.86 mm. For the science beams the lenslets are in the pupil plane, thus the array performs pupil segmentation, creating 4 strips across the pupil. For the tip-tilt beams the array is in the image plane.

Figure 3 shows the two lasers and two continuum sources which can be injected for alignment and calibration. For PAVO alone, one laser would suffice to inject an alignment beam which can be returned by autocollimation at any of a number of points along the beam train, including from the siderostat mirrors themselves. The second laser is included for metrology within the MUSCA system[7] and the continuum sources are used for finding the white-light fringe in MUSCA.

Figure 4 shows an example of a PAVO data frame, which includes both the fringe detection and tip-tilt sensing.

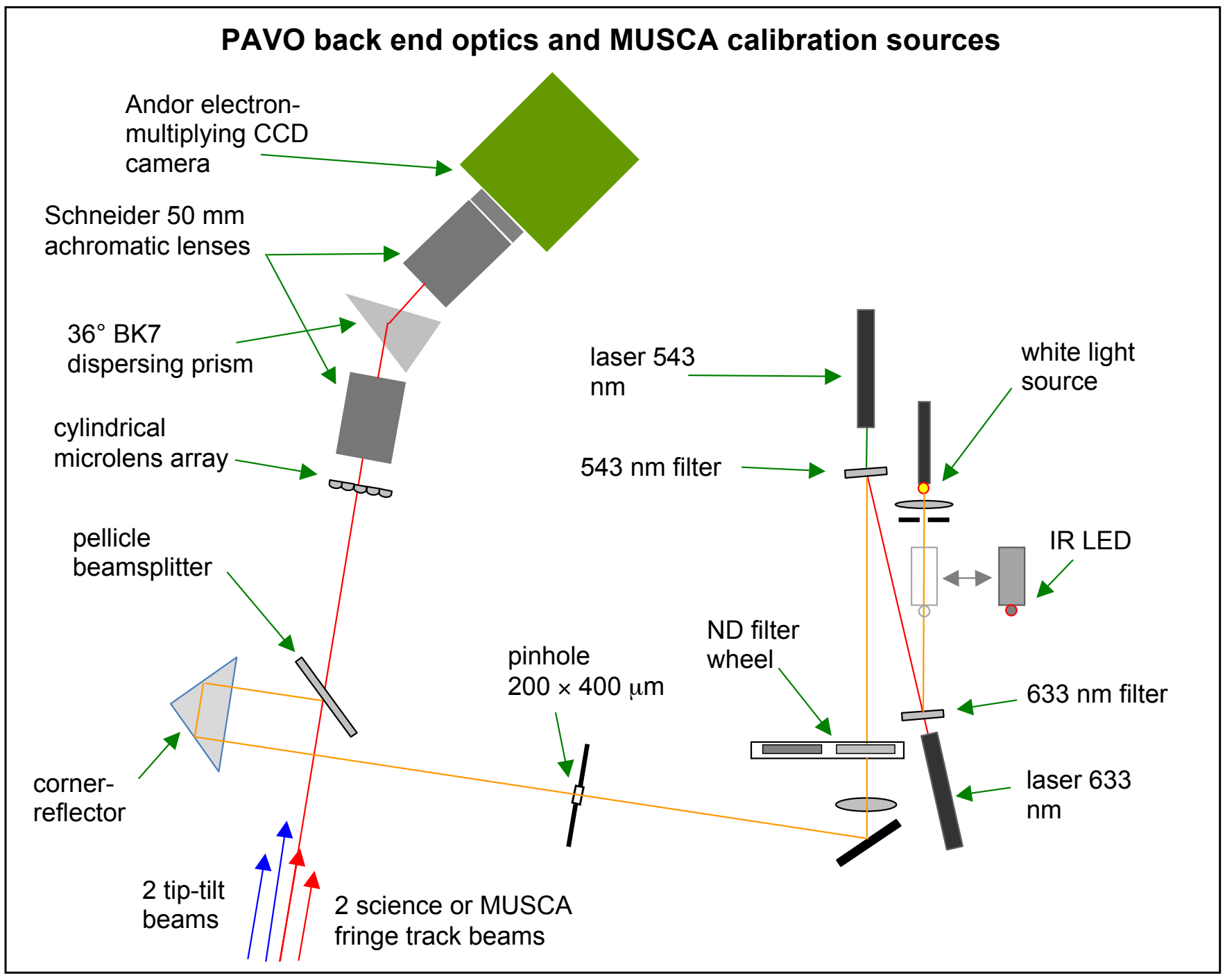

Figure 3. Outline of the light paths and components comprising the PAVO beam combiner and the MUSCA laser and continuum calibration sources.

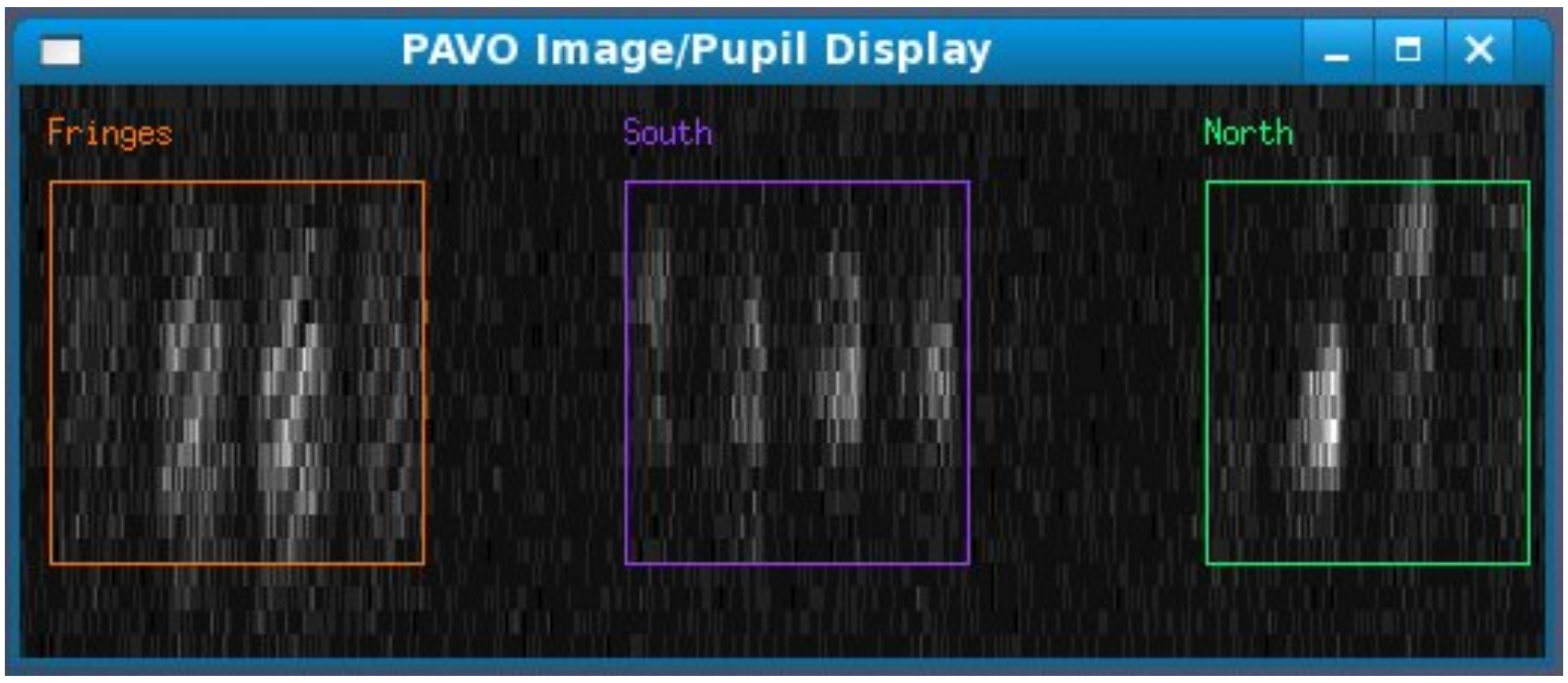

Figure 4. An example of a single 3.5 ms data frame from the PAVO CCD when observing β Cru. In the leftmost of the three boxes the science beams overlap and form fringes. The 4 vertical stripes result from pupil segmentation by 4 lenslets. Within each stripe the horizontal width is due to the spectral dispersion into 24 wavelength channels while the vertical extent represents 24 spatial pixels (binned 8×1) sampling the fringes. The two rightmost boxes in Figure 4 contain the South and North tip-tilt images. In the vertical dimension the position of the star images can be directly determined as the stripes move up and down, while in the horizontal direction the wavelength dispersion is ignored and the image position is found from the balance of intensities among the 4 stripes.

# 3. THE MUSCA DIFFERENTIAL ASTROMETRY SYSTEM

The largest current instrumental development at SUSI is the implementation of the Microarcsecond University of Sydney Companion Astrometry (MUSCA) instrument. The aim of MUSCA is to measure accurately the separations of binary stars in the range 0.1 to 1.0 arcseconds in the case of faint secondary stars and up to 7 arcseconds separation for bright secondary stars ($\alpha$ Cen A,B in particular). The aim is to reach a precision of tens of microarcsec over timescales of several years for the most favourable cases, using observations at 160 m baseline. Monitoring binary separations to this level over a period of years, the perturbations due to Jupiter-mass planets will be detectable to a ratio of separations primary-secondary:star-planet of 10:1.

The principle of MUSCA is to use a scanning mirror to move between the separate fringe packets due to each of the two stars, with precise laser calibration of the scanned interval. While this takes place the fringes will be tracked and locked by simultaneous operation of the PAVO system. Dispersion bias is removed by finding the fringe packet separation as a function of wavelength. The development and status of the MUSCA system are described in a separate paper in this volume[7].

# 4. INSTRUMENTAL UPGRADES

## 4.1 Optical Path Length Compensator (OPLC)

There are two main reasons for carrying out an upgrade and modernization program on the OPLC. Firstly, the current metrology counters and OPLC carriage control hardware use custom electronics based on discrete components, and the servo loop runs on a 25-year old single board computer (AV68K). Lack of spares and the intricacy of the system make this a significant risk to overall reliability.

Secondly, there is a significant loss of fringe visibility at higher carriage speeds. This restricts observations to targets near transit where the path length is changing slowly, and prevents operations on longer baselines. A thorough investigation was undertaken, and it was found that the dominant source of error is a mechanical vibration of the cart in the 10-15 Hz range. An example power spectrum of the vibration is shown in Figure 5. The magnitude of the vibration trends upward with carriage speed. Figure 6 shows the results of a test of carriage position accuracy, as measured by the metrology laser.

The servo loop updates at a rate of 500Hz, so in theory it should be fast enough to compensate for the vibration (with the high speed component of the motion being provided by piezo actuators which drive the piston motion of the OPLC's small secondary mirrors). However, the design of the electronics includes double buffering (to ensure stability of the signals), and the signals go through several transfers between reading the position counters and writing the piezo positions. The updates occur every 2 ms. The result is an 8-10ms delay in the servo loop, which is believed to cause it to lag behind the vibration of the carriage. When the amplitude of the vibration is sufficiently large, the path length changes too much during this lag and fringes are washed out. Thus the second motivation is to increase the speed of the servo loop to update at a rate of approximately 10kHz (actually 9765.625Hz), and reduce the delay to about one cycle of the servo loop (~100μs).

*Method for the upgrade*

The metrology counting circuits were reimplemented in verilog for programming to a FPGA on a PCI hobby board. The circuitry was largely derived from the existing system with a few minor changes. The FPGA was programmed to provide motor pulses and direction signals for the carriage motors as well as the 40MHz reference signal for the metrology (via an onboard oscillator).

A commercial digital to analog converter PCI card is used to provide the analog signals for the piezo actuators of the OPLC secondary mirrors. This card also has sufficient digital inputs to read limit switches, laser status, etc. These two cards reside in a modern PC running a Linux kernel with a realtime patch. The servo loop is in software running on this PC. An additional benefit is that because the servo loop will be running on a modern PC rather than an AV68K with a 20MHz clock, a lot more computing power is available, giving flexibility in modifying the software to carry out diagnostic functions and testing.

*Current Status*

All of the components have been tested independently and they appear to function as intended. Integration and testing of the whole system is underway and the software is nearing completion.

The level of vibration has not yet been reduced. A test was done using rubber isolators, which were very effective in reducing the transmission of higher frequencies from the motor carriage to the optics carriage. However, the isolator worked as a spring creating a 6Hz oscillation that was much larger in amplitude than the problem it was trying to solve. This test established that most of the vibration is coming from the motor carriage. In the future, it would be worthwhile to increase the isolation of the optics carriage from the motor carriage, for example by using a voice coil actuator.

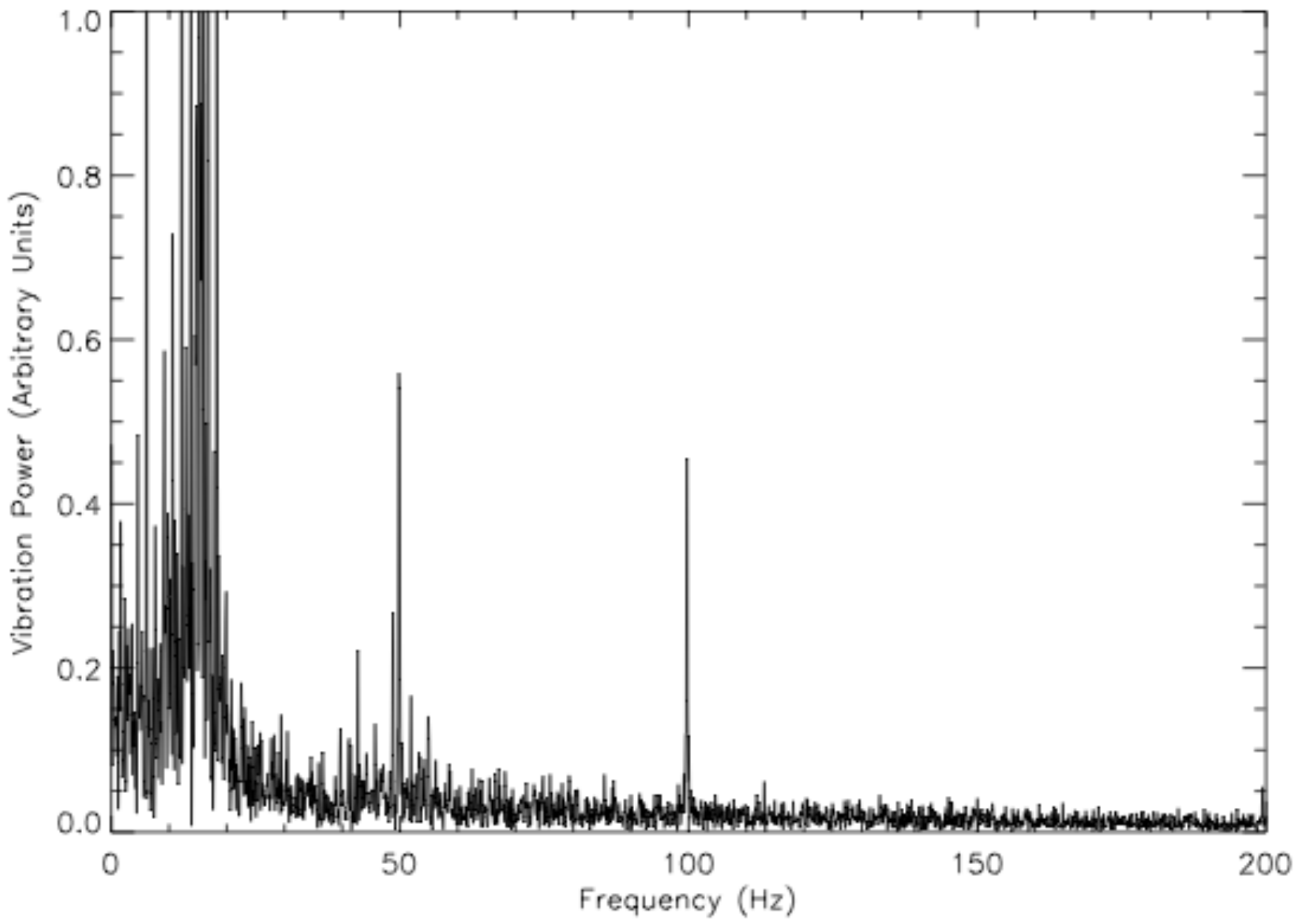

Figure 5. Power spectrum of mechanical vibrations of the OPLC optics carriage when moving at 0.1 mm/s

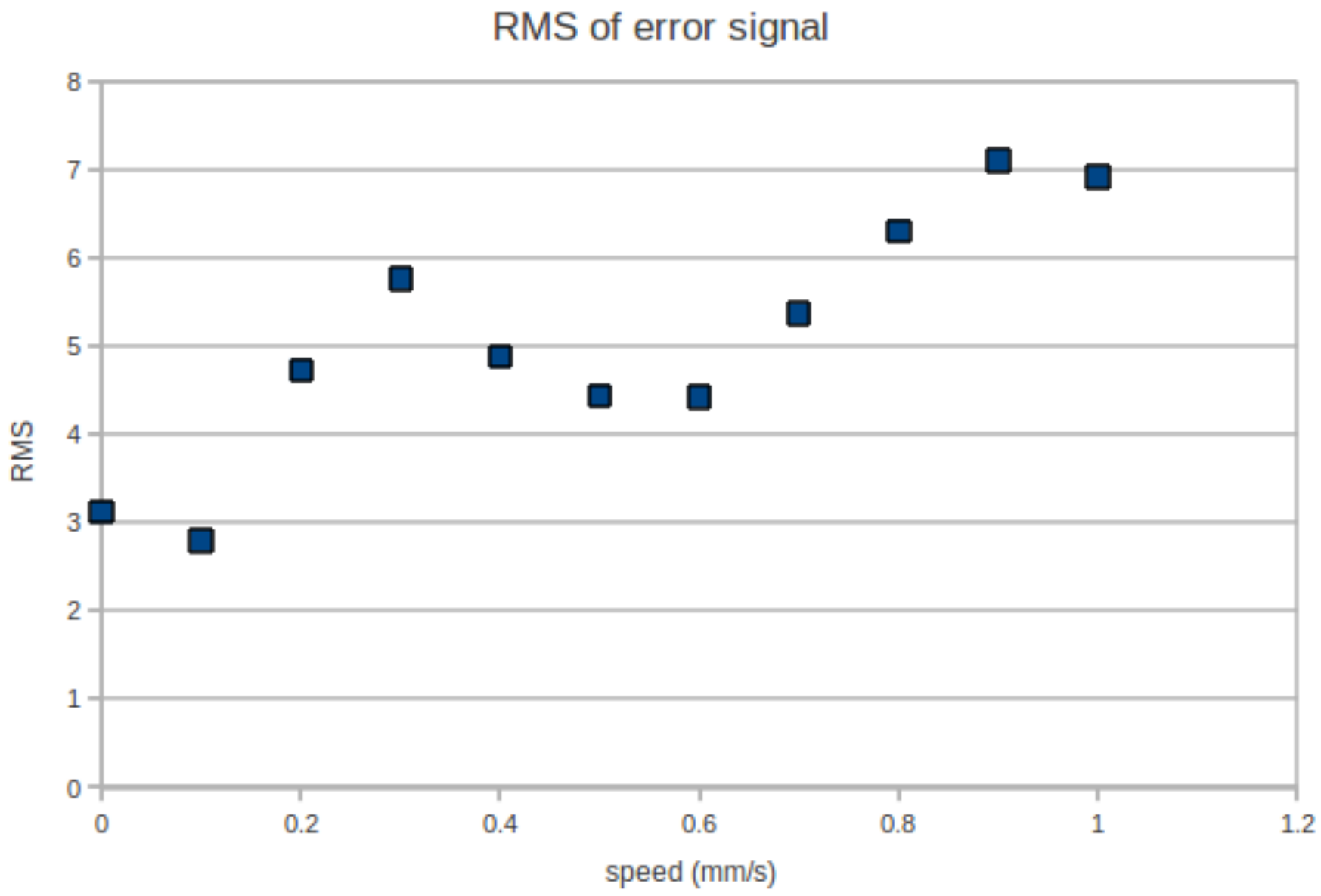

Figure 6. Root-mean-square (RMS) of the carriage position error signal vs carriage speed. The units on the vertical axis are 1/64 of the metrology laser wavelength or ~18nm, thus the full scale 8 units represents 144 nm.

### 4.2 Beam Reducing Telescopes

SUSI as built included a 3:1 beam reducing telescope (BRT) in order to reduce the beam size in the downstream optics from 14 cm to 4.7 cm. It consisted of parabolic mirrors of diameters 546 and 210 mm and focal lengths 1950 and 650 mm respectively. The beams from the north and south siderostats each used a patch on either side of the axis. However, Ireland[8] measured total aberrations equivalent to 0.36λ and 0.28λ P-V surface errors for the portions used by the North and South beams respectively.

The large BRT has now been replaced by two off-axis mirror pairs. Each system consists of a spherical primary of diameter 152 mm and focal length 3.0 m which is used off-axis, and a secondary which is an off-axis hyperboloid of diameter 63 mm and focal length 1.0 m. The secondary is figured to correct the aberrations of the primary. An identical pair is used for each of the North and South beams. Figure 7 shows the new system.

The optical configuration for both the old and new BRT systems is afocal Gregorian. The presence of the internal focus allows a field lens to be placed there in order to increase the field of view (for binary stars), and it is planned to make use of this facility in the replacement BRT.

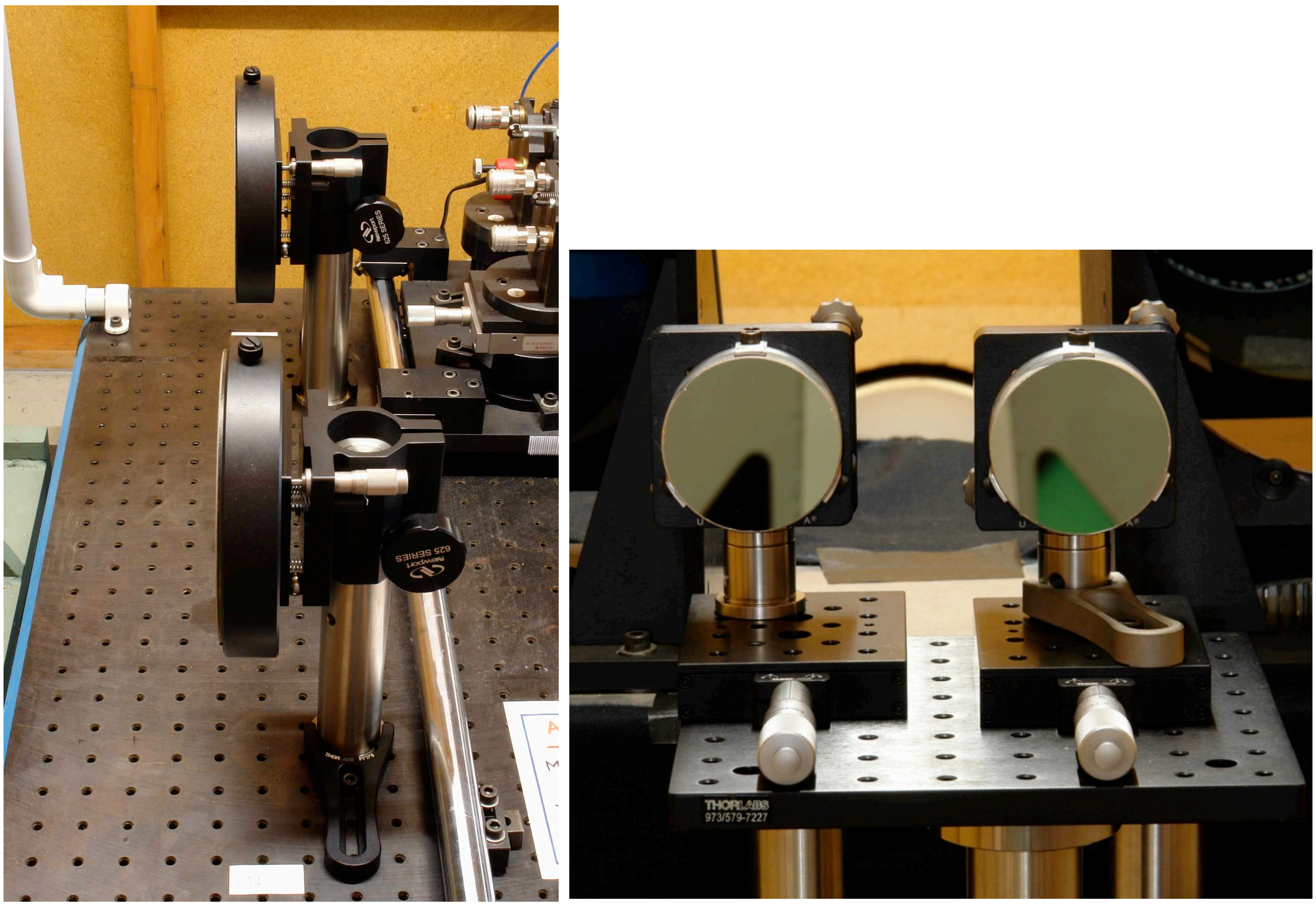

Figure 7. The new beam reducing telescopes at SUSI. Left hand panel: the two 152 mm primary mirrors, seen edge on. Right hand panel: the two 63 mm secondary mirrors.

### 4.3 Local controllers

The original controllers for the siderostat stations and the LDC, which used 68000 processors, have all been replaced with new ones based on off-the-shelf components.[9] Each controller uses a diskless computer designed for industrial process control and two robust digital I/O interfaces that connect to the USB ports on the computer. Four 16-bit digital-to-analog converters provide the drive signals for the siderostat/LDC stepper motors. As well as delivering a modern, low maintenance system the new hardware has been selected to be operable over a wide temperature range.

### 4.4 Infrastructure

Uninterruptible power supplies (UPSs) have been used at SUSI for many years, in order to keep computers running despite the occasional mains power interruptions. More recently, a UPS was installed in each siderostat electronics cabinet in order to enable the siderostat roof to be automatically closed in the event of a power failure while remote observing was in progress. This is critical since a likely cause of mains power failure would be an approaching storm. Unfortunately the summertime temperatures in the electronics cabinets proved too high for the UPS batteries and other components. After numerous UPS failures we have installed a single large (6 KVA) UPS which is located in the air-conditioned tunnel area closest to the siderostats.

### 4.5 Metrology laser

The OPLC carriage location is measured using a stabilized HeNe laser, with an Acousto-Optic Modulator (AOM) in one of two split beams to provide a signal with offset frequency; then the beat frequency between the reference beam and the beam reflected from the carriage provides a measurement of the carriage velocity. This system uses 20-year-old custom optics and electronics, is difficult to maintain and represents a significant failure risk.

As a result, we have begun to investigate the possibility of replacement of this system with a modern one based on a commercially available Fiber Bragg Grating (FBG). An in-fiber AOM is also commercially available. Providing a pump laser is straightforward, and the effects of residual temperature variations on the FBG can be removed by applying controlled strain to the fiber, resulting in an equivalent temperature stability of 0.01°. However, obtaining the required very narrow resonant FBG line width (in order to obtain a sufficiently great coherence length to extend over the full OPLC track) will require investigation.

## 5. SOME SCIENTIFIC RESULTS

### 5.1 Multiplicity Survey of the Sco OB2 Association

Multiplicity properties of recently formed stars can provide valuable insight into star formation mechanisms. The Scorpius-Centaurus-Lupus-Crux OB Association (Sco OB2, Sco-Cen) is the nearest region to the sun with recent massive star formation. In this program all stars within the area of sky occupied by Sco-Cen which were brighter than 5th apparent visual magnitude and bluer than B − V = −0.1 were targeted. The reason for choosing all stars rather than only known Sco-Cen members is that the presence of undetected binarity can affect the HIPPARCOS proper motions upon which the membership assessment was based. Figure 8 shows two example observations, illustrating that the PAVO beam combiner covers a sufficiently large wavelength range that in many cases a binary separation can be found from a single observation.

Of the 69 observed Sco-Cen targets, companions were observed to be associated with 20 of them, 5 of which are new detections. Of the other companions 14 are not well-known or documented systems, none of them have a known orbit and a large fraction do not have a known brightness ratio; hence the observations made in this survey can provide new information on these systems. Full results will be reported elsewhere.

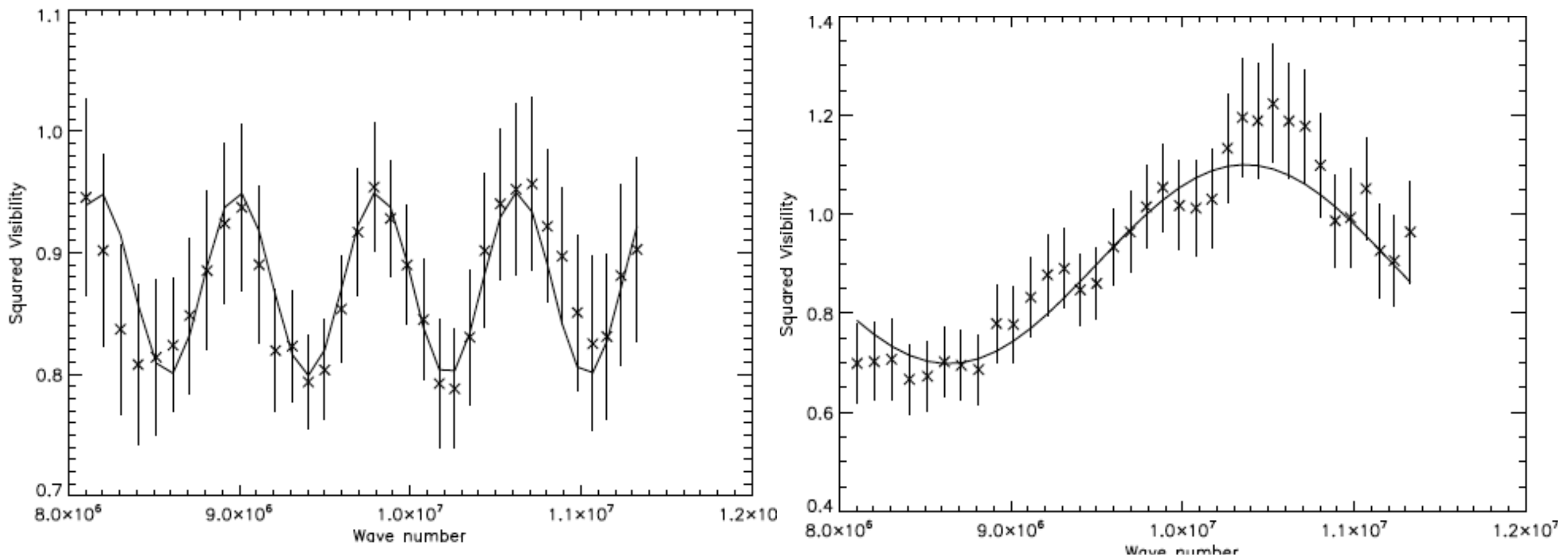

Figure 8. Graphs of $V^2$ vs wavenumber ($2\pi/\lambda$) for observations of ε Cen (left) and $\phi_{02}$ Lup (right). The wavenumber scale covers the PAVO range from λ 810 to 530 nm.

### 5.2 The orbital parameters of κ Sco

The orbital parameters for the double-lined spectroscopic binary system κ Scorpii have been determined from measurements using both the (now decommissioned) red table and PAVO. The orbital parameters were then used in conjunction with available spectroscopic data to determine the masses of the two component stars, as well as the distance to the system using its dynamical parallax. The masses calculated using spectroscopy and Hipparcos data are inconsistent, providing evidence that the parallax or parallax error of κ Sco given in the Hipparcos New Reduction is incorrect. The dynamical parallax is also inconsistent with the Hipparcos parallax, lending further evidence to the claim.

Table 2 gives the orbital parameters, while Figure 9 illustrates the relative orbit. The stellar masses derived from this study are $M_1 = 11.1 \pm 1.7\ M_\odot$ and $M_2 = 10.0 \pm 1.1\ M_\odot$. A full account of this work will be published elsewhere.

Table 2. Orbital parameters of κ Sco, using both the results from Harmanec[10] and allowing all parameters except the period to vary. * indicates fixed parameters, where errors were estimated from Harmanec. d indicates days. T is given as a Julian Date. Note that these results are preliminary.

| | Spectroscopic Results Fixed | | | Fixed Period Only | | |
|---|---|---|---|---|---|---|
| Parameter | Value | | Error | Value | | Error |
| T | 2449984.18 | ± | $0.3^{d}$* | 2449983.23 | ± | $0.4^{d}$ |
| P | 195.458 | ± | $0.007^{d}$* | 195.458 | ± | $0.007^{d}$* |
| a | 10.12 | ± | 0.05 mas | 10.91 | ± | 0.12 mas |
| e | 0.503 | ± | 0.008* | 0.520 | ± | 0.006 |
| $\omega$ | 93.3 | ± | 1.9°* | 87.6 | ± | 0.6° |
| $i$ | 103.27 | ± | 0.09° | 102.3 | ± | 0.2° |
| $\Omega$ | 283.27 | ± | 0.09° | 281.96 | ± | 0.12° |
| $\beta$ | 0.475 | ± | 0.007 | 0.454 | ± | 0.006 |

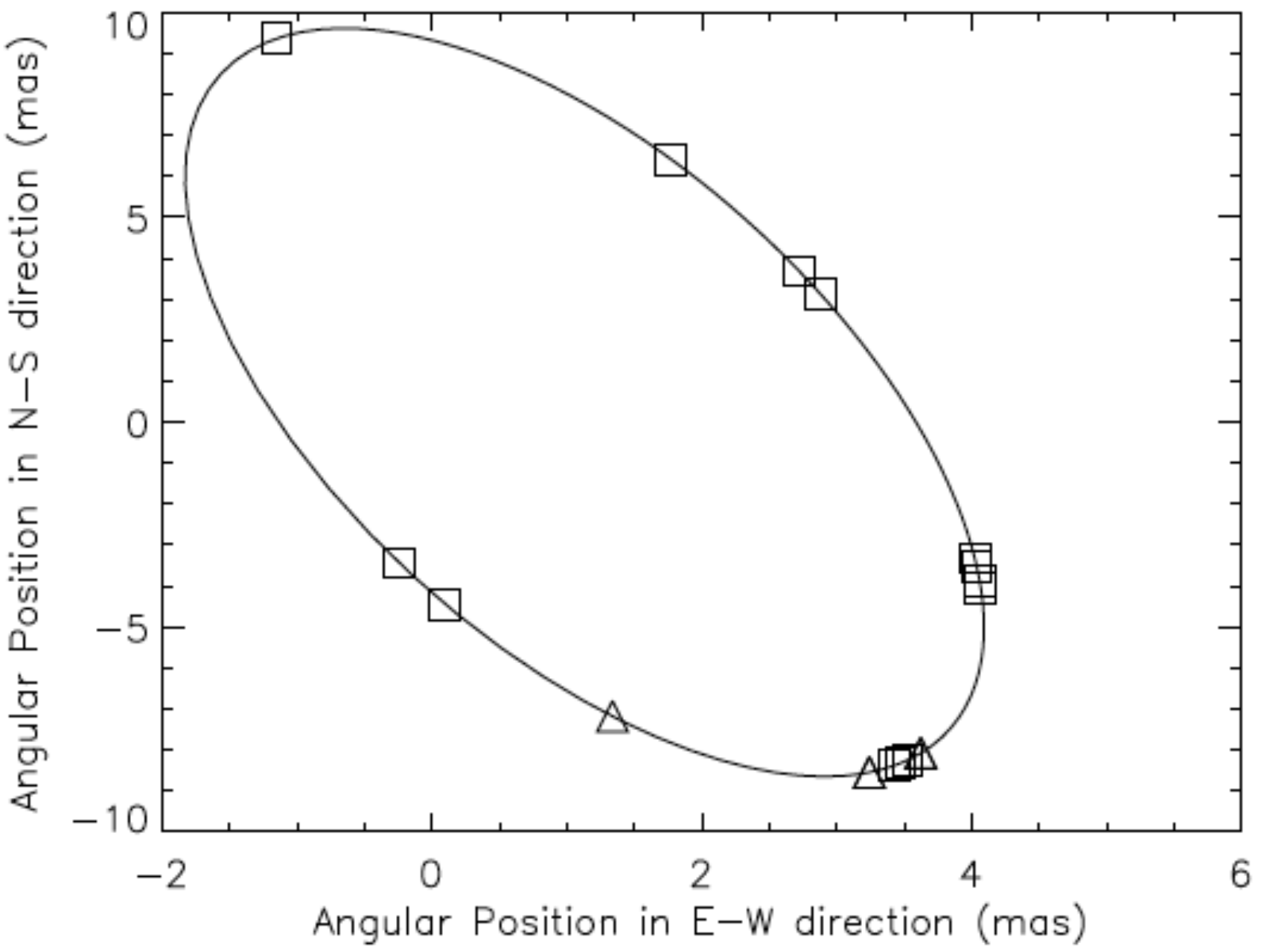

Figure 9. The orbit of κ Sco. Symbols indicate the orbital phase of measurements, but do not represent error boxes.

## ACKNOWLEDGMENTS

This research was supported under Australian Research Council's Discovery Projects funding scheme. Dr. Ireland was supported by an Australian Research Council Postdoctoral Fellowship. The authors would also like to acknowledge the support of the University of Sydney and Macquarie University.

## REFERENCES

[1] Michelson, A. A. and Pease, F. G., "Measurement of the diameter of α Orionis with the interferometer," Ap. J. 53, 249-259 (1921).

[2] Hanbury Brown, R., Davis, J. and Allen, L. R., "The angular diameters of 32 stars," MNRAS 167, 121-136 (1974).

[3] Davis, J. and Tango, W. J., "New determination of the angular diameter of Sirius," Nature 323, 234-235 (1986).

[4] Davis, J., Tango, W. J., Booth, A. J., ten Brummelaar, T. A., Minard, R. A. and Owens, S. M., "The Sydney University Stellar Interferometer I. The instrument," MNRAS 303, 773-782 (1999) .

[5] Davis, J., Ireland, M. J., Chow, J., Jacob, A. P., Lucas, R. E., North, J. R., O'Byrne, J. W., Owens, S. M., Robertson, J. G., Seneta, E. B., Tango, W. J. and Tuthill, P. G., "The Sydney University Stellar Interferometer: A Major Upgrade to Spectral Coverage and Performance," Pub. Astron. Soc. Aust., 24, 138–150, (2007).

[6] Ireland, M. J., Mérand, A., ten Brummelaar, T. A. Tuthill, P. G., Schaefer, G. H., Turner, N. H., Sturmann, J., Sturmann, L. and McAlister, H. A., "Sensitive visible interferometry with PAVO," Proc. SPIE 7013, 701324 (2008).

[7] Kok, Y., Ireland, M.J., Tuthill, P.G., Robertson, J.G., Warrington, B.A. and Tango, W.J., "Self-phase-referencing interferometry with SUSI, " Proc. SPIE 8445-72 (these proceedings) (2012).

[8] Ireland, M. J., "Optical Interferometry and Mira variable stars," Ph.D. thesis, University of Sydney (2005).

[9] Tango, W. J. and Ireland, M. J., "A new embedded control system for SUSI," Proc. SPIE 7734-140 (2010).

[10] Harmanec, P., Uytterhoeven, K. and Aerts, C., "Disentangling component spectra of κ Sco, a spectroscopic binary with a pulsating primary I," A&A 422, 1013–1021 (2004).